\documentclass[12pt]{article}
\usepackage{graphicx}
\usepackage[top=3cm, bottom=2cm, left=3cm, right=3cm]{geometry}
\usepackage{amssymb}

\title{Ultra-high energy Cosmic Rays and the Extragalactic Gamma Ray Flux.}
\author{$^1$A.D. Erlykin $^{a,b}$ and A.W. Wolfendale $^{b}$\\
$(a)$ P N Lebedev Physical Institute, Moscow, Russia.\\
$(b)$ Physics Department, Durham University,\\ Durham, DH1 3LE, UK}
\date{\today}

\begin{document}
\maketitle

\begin{abstract}
Ultra-high energy cosmic rays interacting with the radiation fields in the universe
cause electromagnetic cascades resulting in a flux of extragalactic gamma rays,
detectable to some 100  GeV. Recent precise measurements of the extragalactic gamma ray
 flux by Fermi-LAT, coupled with estimates of the background from active galactic
nuclei of various types, allows limits to be set on the cascade component. By
comparison with prediction and, making various assumptions, ie taking a particular
model, limits can be set on the maximum energy to which ultra-high
 energy particles can be accelerated.

If our model is correct, it is unlikely that the maximum energy is above 100 EeV, in
turn, the apparent `GZK' cut-off in the measured ultra-high energy spectrum could
instead be due to a fall-off in the intrinsic emergent particle spectrum. However, it
is not possible to be dogmatic at the present time because of uncertainty in many of 
the parameters involved. We have used recent estimates of the range of parameters and 
have found that although our model has parameters in the allowable ranges the 
uncertainties are so large that our result is not unique, although the method is 
satisfactory. The result must thus, so far, be taken as an indication only.
\end{abstract}
\footnote{Corresponding author: tel +74991358737 \\
 E-mail address: erlykin@sci.lebedev.ru}
\section{Introduction}
Some decades ago [1] it was pointed out that there is a constraint on the energy 
spectrum of the ultra-high energy cosmic ray (UHECR) intensity caused by the 
electromagnetic cascade of photons and electrons initiated by the initial interaction 
of the UHECR with the cosmic microwave background (CMB). Later interactions with both 
the CMB and the starlight and infra-red fields cause the cascade to extend down to the 
MeV gamma ray region.

Comparison of the expected flux for various (proton) production scenarios [2] with the 
then measured extragalactic (EG) gamma flux [3] allowed limits to be put on the former.
 With more recent estimates of the EG gamma flux by Fermi-LAT [4] and superior 
estimates of the contribution from unresolved sources (active galactic nuclei, AGN) 
[5] an improved production scenario can be derived.  Our particular interest is the 
maximum energy achievable at the sites of acceleration, the analysis being made for 
both primary protons and iron nuclei. Regarding the mass composition at the highest 
energies, the case for heavy nuclei was made early on (see, eg, [6]) and recent 
analysis give some support [7,8,9].

In what follows, we start by assuming that our initial calculations [1,2] were correct 
and follow them through with the Fermi-LAT gamma ray spectrum [4] to establish the 
method and arrive at tentative conclusions. This is followed by an examination of the 
limits that can be put on the various input parameters, and on the consequent result 
using a recent analysis [9].

Of relevance is the observation by the Pierre Auger Observatory (PAO) that above
55 EeV, some 24\% of the particles come from the giant Cen-A radiogalaxy within 
a direction of 40 degrees [10] to a significance level of 4\%. Others [11,12] have 
developed theories assuming this identification. The relation of this result to the 
problem in hand will also be examined.

\section{The diffuse EG gamma ray spectrum.}
Figure 1 shows the Fermi-LAT spectrum [4] in comparison with our earlier estimate [13].
 The steeper contemporary spectrum can be understood in terms of a more accurate
'discrete source' correction, the new measurements allowing weaker sources to be
identified. The effect of the recent [5] correction for AGN is shown. It will be noted
that the EG gamma ray flux available for cascade gamma rays is small, ie of order of
the discrete source contribution itself.
\begin{figure}[ht]
\begin{center}
\includegraphics[width=15cm,height=9cm]{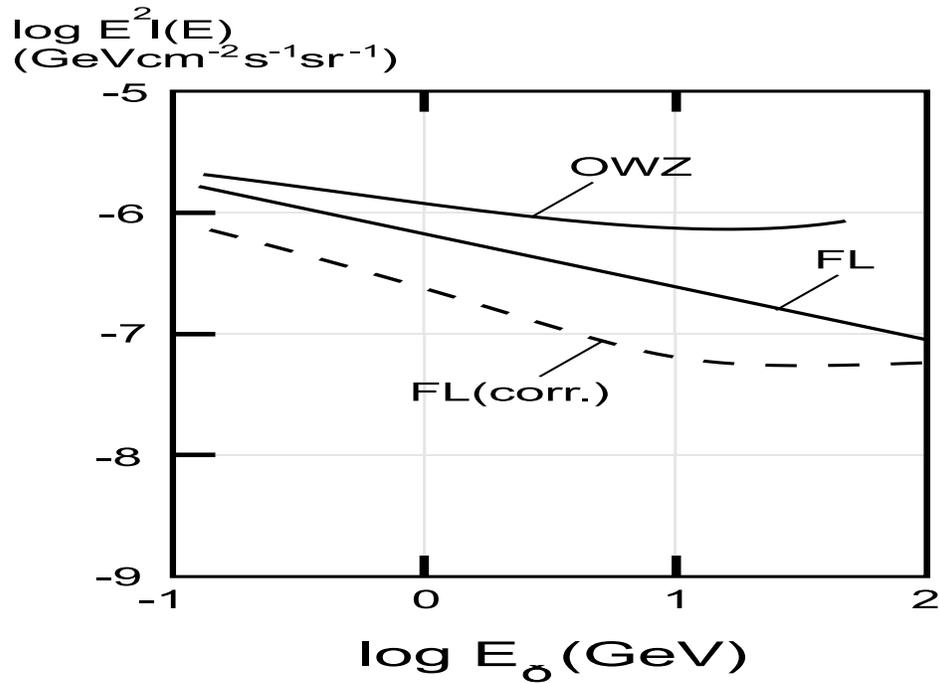}
\end{center}
\caption{\footnotesize Comparison of the Fermi-LAT EG diffuse gamma ray spectrum [4],
denoted FL with that derived earlier by us [13] and denoted `OWZ'. `FL(corr)' relates
to the Fermi-LAT spectrum corrected for AGN.}
\label{fig:fig1}
\end{figure}
\section{The cascade gamma ray spectrum.}
\subsection{Effect of maximum energy and mass.}
The cascade calculations referred to in [2] involve determining the energy abstracted
from the EG particle beam by the interactions with the CMB, it is this energy that is
propagated by way of subsequent interactions with the radiation fields to give a gamma 
ray spectrum that extends down to the MeV region. The intensity in the GeV region is 
proportional to the energy abstracted. This latter is derived in a straightforward way 
from application of the attenuations to the assumed primary (injection) spectrum.

The profile of the attenuation mean free path $(\lambda)$ versus energy for both
protons and iron nuclei has been known for many decades. Using recent calculations
[14,15], for protons $\lambda$ falls from its red-shift value of 4000 Mpc to 1000 Mpc 
at 10 EeV, 200 Mpc at 100 EeV and to 14 Mpc at 1000 EeV. Corresponding values for Fe 
are: 3200 Mpc at 10 EeV, 630 Mpc at 100 Eev and about 2 Mpc at 1000 EeV. The results in
 [2] were for protons with an injection spectrum of E$^{-2}$ and a maximum energy of 
1000 EeV. The relative gamma ray intensities are given in Figure 2.
\begin{figure}[hbtp]
\begin{center}
\includegraphics[width=10cm,height=7cm]{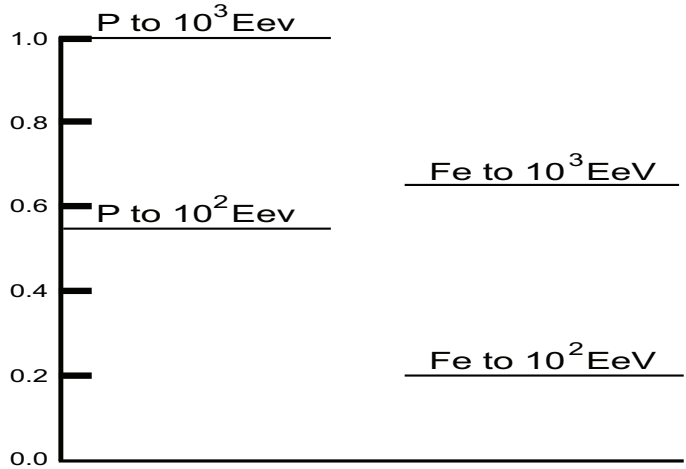}
\end{center}
\caption{\footnotesize Contributions to the EG diffuse gamma ray intensity in the GeV
region for two masses: protons and iron and two maximum injection energies: 10$^2$ and
10$^3$ EeV. Protons to a cut-off at 10$^3$ EeV' is taken as the datum. The ordinate is 
the relative contribution normalised to the standard of protons to $10^3$ EeV.}
\label{fig:fig2}
\end{figure}
\subsection{Sensitivity to the evolutionary model and other parameters.}
In [2] we varied the (proton) injection spectrum with energy and maximum red-shift 
(z$_m$) using our earlier analysis as G(E,z)dz = BE$^{-\gamma}$f(z) dz, where \\
f(z) = H$_0$$^{-1}$ (1+z)$^{\beta-5/2}$ for z $<$ z$_m$ and f(z) = 0 for z $>$ z$_m$.
H$_0$ is the Hubble constant.

Figure 3 (from [2]) gives the predicted spectra below 0.1 PeV for 3 variants of the
parameters:\\
NC - no cosmological increase\\
C1 - $\beta$ = 3.7, z$_m$ = 4\\
C2 - $\beta$ = 3.7, z$_m$ = 9\\
In [2], C1 was regarded as the standard, ie for `normal galaxies' an injection
intensity varying as (1+z)$^{1.2}$ with z$_m$ = 4 ( this value approximating the
redshift at which the volume density of active galactic nuclei begins to fall ). The
energy content: 1.3x10$^{-7}$ eVcm$^{-3}$ [2] is close to the value calculated from
the assumption of an E$^{-2}$ EG injection spectrum normalised to the contemporary
measured spectrum [10] at 10 EeV.

Turning to the radiation fields, later work [14], has shown that the IR intensity is a 
factor 2 lower than assumed in [2]. However, the IR affects the ensuing gamma ray 
spectrum only above about $10^3$ GeV and thus at higher energies than of interest here.
The transition between Galactic and Extragalactic UHECR is still the subject of debate.
Here we adopt 2 EeV as in all our previous and contemporary work. The use of a lower 
cut-off is not large because the initial gamma-particle interactions only start in the 
EeV region. What is important is the intensity and slope of the particle spectrum and 
another EG spectrum can easily be introduced into the calculations. 
\begin{figure}[ht]
\begin{center}
\includegraphics[width=15cm,height=10cm]{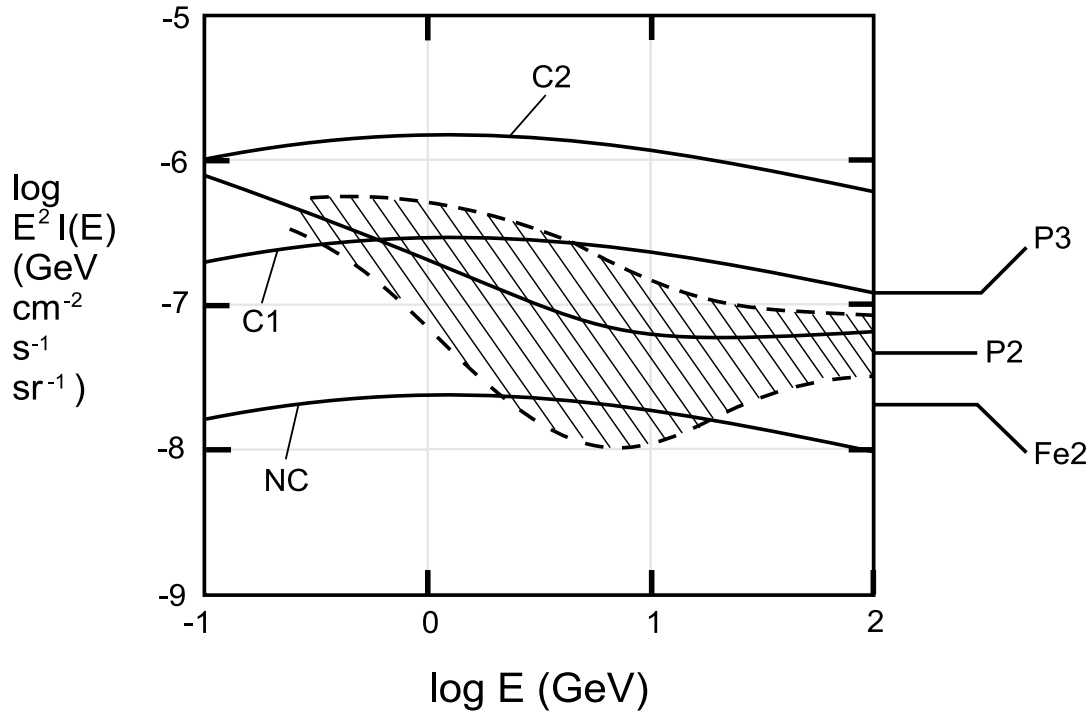}
\end{center}
\caption{\footnotesize Comparison of the predicted EG diffuse gamma ray spectrum from
[2] for:
\newline
NC, no cosmological increase in injection,\newline
C1, the datum: injection as (1+z)$^{\beta-5/2}$, with $\beta = 3.7$ and 
Z$_m$ = 4
\newline C2, enhanced cosmological injection, with $\beta$=3.7 and z$_m$ = 9.
\newline The spectra are approximate. P2, P3 and Fe2 relate to the expected 
contributions from
cascading from Figure 2. P2, P3 relate to protons with maximum energies 10$^2$ and
10$^3$ Eev, and Fe2 is for iron, limited to 10$^2$ EeV. The shaded area indicates the
spread in estimates of the diffuse flux.
In view of the many uncertainties in the values for the parameters ( IG radiation
fields, magnetic fields, primary CR spectra ) the values for Fe2, P2 and P3 must be
regarded as illustrative only.}
\label{fig:fig3}
\end{figure}
\section{The maximum CR energy.}
\subsection{The analysis of EG gamma rays.}
Inspection of figure 3 shows an anomaly in that the shaded region rises outside the 
expected range for E$_\gamma\lesssim$ 3 GeV; this is a consequence of the observed 
gamma ray spectrum having a slope near -2.4, in comparison with the expected spectra 
having slopes near -2.0.

The reason for the discrepancy is probably the effect of CR electrons escaping from
normal galaxies and interacting with the ambient radiation fields via the Inverse
Compton effect. For such normal galaxies with an escaping electron flux above about 100
 GeV having a slope -4.0, the eventual gamma ray spectrum would have slope $\sim$ -2.5,
 as required. Most of the electron-gamma rays would come from CR produced at high red
shifts.

The contribution of the above from above 10 GeV would be expected to be small because
of the likely dearth of electrons of sufficiently high energies and thus we just adopt
the gamma ray energy range in Figure 3 of 10 to 100 GeV.

The energy content available for the electromagnetic cascade is an obvious indicator of
 the particle charge, maximum energy combination, assuming, as usual, that the EG CR 
injection spectrum has a slope $\sim$ -2. Figure 3 shows the various possibilities
normalised to the NC (p, 1000 EeV) results. ( Note the proviso.)

Although the sensitivity of the flux to the combinations is not large (in terms of the 
spread in estimates of the diffuse flux - shown shaded) it is evident that predictions 
are in `the right region'. Specifically, taken literally, and using the results for 
E$\gamma$ : 10-100GeV, and following the PAO indications that the particles above 30 
EeV are of `mixed composition' [11,12], the results appear to indicate a maximum energy
 of about 100 EeV.

At this stage mention can be made of later work on this topic [16]. Understandably, the
 results are not identical, the differences being largely due to differences in
essentially all the values for the many relevant parameters: the intergalactic
radiation fields and magnetic fields, the EG CR spectrum and its mass
composition, the most likely form of the injection spectrum: its $\beta$ and $z_m$
values. The significance of these later results will be considered in $\S$5.
\subsection{The PAO results for Centaurus-A}
There is rather strong evidence that some, at least, of the UHECR come from the
nearest, very strong, radiogalaxy, Centaurus-A. Thus, PAO finds that about 24\% of the
particles can have come from this object, when allowance is made for the spread in
arrival directions caused by the intergalactic and interstellar magnetic fields [10].

Figure 4 shows the PAO energy spectrum. Insofar as Cen-A is only some 2-3 Mpc away, the
 losses on the CMB are quite negligible and the true injection spectrum should be
observed. With a slope of -2 the Cen-A spectrum should look as indicated. It is evident
 that the particles expected above 100 Eev are absent. Thus, this source of UHECR, at
least, which would have been expected as a for-runner in the search for UHECR sources,
 does not emit particles above 100 EeV.
\begin{figure}[h]
\begin{center}
\includegraphics[width=10cm,height=8.9cm]{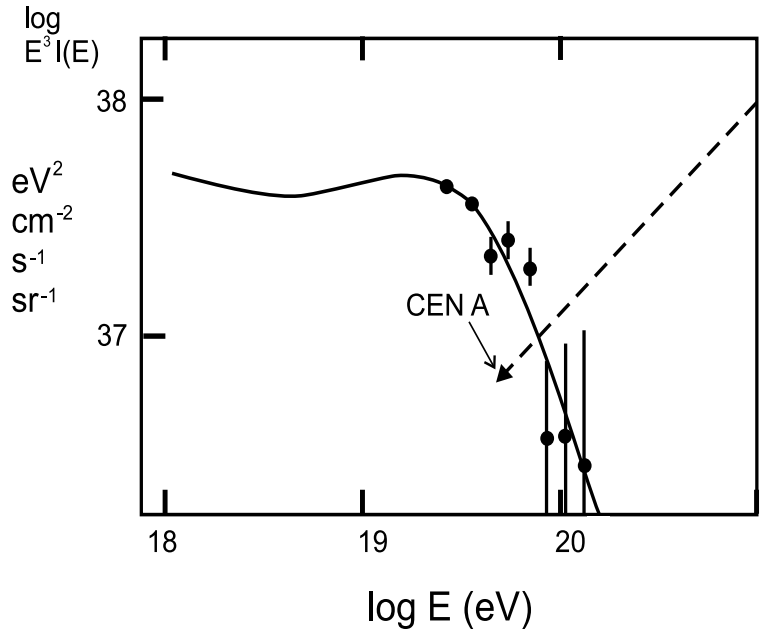}
\end{center}
\caption{\footnotesize The PAO spectrum [10] and our estimate of what the Cen-A
spectrum would look like if it were typical for the strongest EG sources and had
spectral shape $\sim E^{-2.0}$ at injection and continued to 10$^{21}$ eV. Insofar as 
the identification of Cen-A as the source of UHECR ( 4\% probability of no effect [10])
 the arguments are not (yet) firmly based.}
\label{fig:fig4}
\end{figure}

\subsection{Further work needed.}
There are two other areas where further work is needed. The first concerns the measured
 UHECR spectrum from the PAO. This observatory is in the Southern Hemisphere and there
is the possibility that the measured spectrum is not representative of the Global
average. This follows from the likely non-uniform volume density of the sources over
the Universe. That the volume density of `normal' galaxies is not uniform has been
known for some time. Recently, [17], the presence of a North-South anisotropy has been
confirmed ( at least for galaxies in the so-called N and S polar caps ), the mean ratio
 of the N/S densities varying with distance, d, as: 4.3
locally, 3.5 at d = 25 Mpc, 1.6 at d = 50 Mpc, 1.23 at d = 75 Mpc, 1.33 at d = 100 Mpc
and $\sim$1.0 for d: 100-300 Mpc. Thus, if UHECR sources are distributed in the same
way as 'galaxies' from the survey , there should be a difference between the UHECR
intensities at such high energies where diffusion has not diluted the N-S anisotropy
too seriously. It should be noted that the effect of a spectrum of primary masses will
reduce the observed anisotropies if, as is likely, the anisotropy ( amplitude and phase
 ) is rigidity dependent.

It is interesting to note the increase in intensity of the EG component above 10$^{19}$
eV with increasing energy for N with respect to S detectors [18], suggestive of
the above hypothesis. It is possible that the well known N,S
difference in apparent mass compositions above 10$^{18}$ eV is also
related to the N,S densities: the excess of nearby galaxies in the
North will lead to an excess of light nuclei (~He, CNO, but not
protons~) because of their short attenuation lengths. For example, for
E$>6\cdot 10^{19}$ eV, the attenuation length is only 20 Mpc for CNO, whereas it is
approaching 100 Mpc for protons and iron [18]. However, it is impossible to be dogmatic
 at this stage because the measured galaxy densities versus distance are only valid for
 restricted regions of the Universe and there is the ubiquitous problem of the effect
of both Extragalactic and Galactic magnetic fields in smearing the arrival directions.
 The relevance of the above to the problem in hand is that there is still uncertainty
in the effective average UHECR intensity, on a local cosmic scale, to use in the
calculations and, correspondingly, the mass  composition to take.

The second area relates to the values of the many parameters used in the calculations:
$\beta$ and $z_m$ (Figure 3), the IG radiation and magnetic fields, the CR spectra and
their masses and so on. An example is of z$_m$ in the cosmological increase in CR
injection efficiency. Recent work for AGN (eg [19]) suggests that it might be somewhat
higher than the value of 4 adopted. If so, the constraint on the maximum CR energy
could be more stringent.

\section{More recent estimates of the EG gamma ray flux.}
As mentioned earlier, recent analyses have been made [16] which are similar to ours [2]
 but using Fermi-LAT data for the observed gamma ray spectrum and adopting somewhat 
different values for some of the important parameters. For example, whereas we use an 
exponent 2.0 (the Fermi-value) for the true EG spectrum the workers in [16] use 
$\gamma$: 2.3 to 2.6. Another example in the value of z$_{max}$ (see $\S$3.2); we used 
Z$_m$=4, and justified it, whereas [16] adopted z$_{max}$ = 2. The calculations in [16]
 were for protons only, clearly at variance with the experimental situation [7, 8, 9].

The aim of [16] was to find the best-fit values of `n' (related to our $\beta$) and the
 exponent of the EG spectrum, $\gamma$. The `HighRes' spectrum [20] was used to derive 
the EG spectrum, with its required $\gamma$-value, to give a best-fit to the measured 
Fermi-LAT EG diffuse gamma ray spectrum. Comparison can be made via the derived energy 
content in the predicted gamma ray spectrum. Our value was $20\cdot 10^{-6}$eVcm$^{-3}$
, of values in the range 1-6 eVcm$^{-3}$ in [16]. The difference in values between 
ourselves and [16] can be explained to within $\pm$30\% by the differences in $\gamma$,
 n($\beta$), input CR proton spectrum, etc. 

What the computed energies do bring home is the sensitivity of the outcome (eg 
predictions of Figure 3) to the values for the parameters referred to, some of which 
are still poorly known. For example, changing the predictions by a factor 2 (not 
impossible) would change the conclusions considerably: for example, from P2 to Fe2.

\section{Neutrino fluxes}
The IceCube experiment [21] has recently observed high energy events indicative
of the presence of neutrinos of energy in the PeV region: 2 events here and over 10
back to a TeV. It is interesting to note that back in the 1960s the KGF experiment [22]
observed one such event - the 'golden event' of very high multiplicity. Such events are
 presumably due to neutrinos from the initial primary - CMB interaction at cosmological
 distances and the neutrinos, unlike the gamma rays, do not have a 'threshold radius'.
Other things being equal, the median distance of origin will be half the Hubble radius,
 i.e. of order 2000 Mpc compared with the gamma rays of main concern here which have a
median distance of only 5-70 Mpc. In the calculations [eg 20] the assumption has been
immediately made that the UHECR spectrum is uniform throughout the Universe ( apart
from its z-dependence ). This need not be so. The discussion in \S4.3 regarding the
volume density of galaxies, which varies by at least $\pm 2$ over the local 100 Mpc
means that there is at least a further factor of $\pm 2$ uncertainty in the predicted
neutrino intensities, this is an addition to the uncertainty in the primary mass and
maximum energy of the very distant UHECR.

The moral of the neutrino analysis is that differences of significance between observed
 and expected TeV to PeV intensities need not indicate exotic processes; more
conventional differences need to be dismissed first.
\section{Discussion and Conclusions}
The very tentative result of the present work is that there is no evidence of UHECR
being emitted from sources with energy above 100 EeV. More accurate estimates of the
electron contribution to the EG diffuse gamma ray intensity are needed, as are further
measurements of the UHECR spectrum and astronomical evidence for the spatial
distribution of likely CR sources on 100 Mpc scales. Furthermore, values for the many
parameters in the model need to be made. Indirect studies by way of high energy CR
neutrino fluxes will also be of great value.

\vspace{5mm}

\large{\bf Acknowledgments}

The authors are grateful to the Kohn Foundation for financial support.

\vspace{1cm}

\large{\bf {References}}
\begin{enumerate}
\item Wdowczyk, J., et al., 1972, J.Phys. A. 5, 1419.
\item Wdowczyk, J. and Wolfendale, A.W., 1990, Astrophys. J. 349, 35.
\item Fichtel, C.E. et al., 1978, Astrophys.J., 222, 833.
\item Abdo, A. et al., 2010, Phys. Rev. Lett., 104, 101101.
\item Abazajian, K.N. et al., 2011, Phys. Rev. D 84,103007; arXiv:1012.1247.
\item Tkaczyk, W . et al., 1975, J.Phys.A. 8, 1518.
\item Wibig, T. and Wolfendale, A.W., 2009, Open Astron. Journ. 2, 95.
\item Abraham, J. et al., 2010, Phys. Rev. Lett., 104, 091101.
\item Watson, A.A., 2014, Rep. Progr. Phys., 77, 036901(24pp); arxiv:1310.0325
\item Abreu, P. et al. for the Pierre Auger Coll., 2010, Astropart. Phys., 34, 314;
      arxiv:1009.1855
\item Suchov, O.B. et al., 2012, Adv. in Astron. and Space Phys. 2, 73.
\item Kim, H.B., 2013, J. Korean Phys. Soc. 62, 708.
\item Osborne, J.L. et al., 1994, J.Phys. G. 20, 1089.
\item Ord, M. et al., 2011, 32nd Int. Cosm. Ray Conf., Beijing, 8, 117. 
\item Hooper, D. et al., 2007, Astropart. Phys., 27, 199; astro-ph/0608085.
\item Ahlers, M. et al., 2010, Astropart. Phys., 34, 106; arxiv:1005.2620
\item Whitbourn, J.R. and Shanks, T., 2013, Mon. Not. Roy. Astron. Soc.;
arxiv:1307.4405.
\item Watson, A.A., 2013, IACTPP Villa Olmo Conf.: \\
      $http://villaolmo.mib.infn.it/presentations/IACTPP\_2013$
\item Fontanot, F, et al., 2012, Mon. Not. Roy. Astron. Soc. 425, 1413.
\item Aartsen, M.G. et al., 2013, Phys. Rev. Lett., 111, 021103; \\ 
      2013, Science, 342, 1242856; arxiv:1304.5356; arxiv:1311.5238.
\item Ahlers, M. and Halzen, F., 2012, arxiv:1208.4181.
\item Achar, C.V. et al., 1965, Proc.9th Int. Cosm. Ray Conf., London, 2, 1012.

\end{enumerate}

\end{document}